# True 3D imaging with monocular cues using holographic stereography

Yi-Ying Pu, Bing-Chu Chen, Yuan-Zhi Liu, Jian-Wen Dong*, and He-Zhou Wang**

*State Key Laboratory of Optoelectronic Materials and Technologies,*

*Sun Yat-sen ( Zhongshan ) University, 510275, Guangzhou, China*

* *dongjwen@mail.sysu.edu.cn*, ** *stswhz@mail.sysu.edu.cn*

**Abstract**

A quantitative condition is derived to evaluate the monocular accommodation in holographic stereograms. We find that the reconstruction can be viewed as true-3D image when the whole scene is located in the monocular cues area, with compatible monocular cues and binocular cues. In contrast, it reveals incorrect monocular cues in the visible multi-imaging area and the lacking information area. To demonstrate our theoretical predictions, a pupil-function integral imaging algorithm is developed to simulate the mono-eye observation, and a holographic printing system is set up to fabricate the full-parallax holographic stereogram. Both simulation and experimental results match our theoretical predictions.

Holographic stereography [1,2], which is a hybrid of holography and integral photography [3-5], is one of the promising technologies for three-dimensional (3D) displays [6]. Considered as a convenient way to record dynamic 3D objects, holographic stereography has attracted extensive interest in the past decades [7,8]. Nevertheless, it is controversial whether the holographic stereogram (HS) is true 3D imaging because it includes the electromagnetic phase related to propagating directions, but excludes the depth and neighboring points on the hologram [9,10].

Human beings perceive 3D images using both physiological and psychological cues. Psychological cues include monocular cues (MC), motion parallax, binocular cues, and vergence cues. Among them, binocular cues and MC cues are both principal functions of real-life 3D views. For the absence of some phase information, nowadays the HS mainly gives binocular cues and incorrect MC. This discrepancy between them causes visual fatigue. Recently, several qualitative strategies have been proposed to evaluate the MC in multi-view displays [9-12]. For example, a group from Japan proposed the phase-added stereograms that converged to Fresnel holograms [10] and a MIT group generated their panoramagram using controllable "wafels" [12]. However, the quantitative studies and experimental demonstrations are still needed to clarify the MC in the HS.

In this Letter, three categories of areas are distinguished for the first time, namely, the MC area, the visible multi-imaging (VMI) area, and the lacking information (LI) area. We find that the correct MC occurs when the whole scene falls into the MC area. The theoretical predictions have been validated by the simulations using the pupil-function integral imaging algorithm, as well as the full-parallax HS with correct MC fabricated by the HS printing system.

In a point source hologram, a sharp point image is observed when point P is focused [see Fig. 1(a)]. Otherwise, a blurry spot is observed. In this way, the depth of the point is estimated by the mono-eye. In the HS, the wave radiated by the point source passes through a lens and then a beam-width limit pinhole. Thus the wave can be viewed as a geometrical ray. In this way, the image point has the same size as the holographic element (hogel) when the mono-eye focuses correctly [see Fig. 1(b)]. Because the hogels are discrete, multiple spots can be observed when the eye focuses beyond the scene point. It confuses the viewer with the depth. However, when the distance of two adjacent spots is within the eye's resolution, these spots become a big blurry spot, which leads the depth of field being perceived.

Next a quantitative condition is derived to evaluate the MC. Consider two adjacent hogels nearest to the scene point P [see Fig. 1(c)], we have

$$\theta = 2\arctan\frac{\gamma|z_P - z_F| - h}{2z_F} \qquad (1)$$

Here $\gamma = h/|d_H - z_P|$, $h$ is the size of the hogel, $z_F$, $z_P$ and $d_H$ are the distances from the eye to the focal plane, point P, and the HS, respectively. Note that Eq. (1) is still applicable even when point P is behind the HS. When two adjacent points are undistinguishable, the angle $\theta$ is less than the lateral resolution of the naked eye (1.5' as usual). Hence, the focus area is determined by

$$\frac{z_P\gamma/2 - 1}{\gamma/2 + \tan 0.75'} < z_F < \frac{z_P\gamma/2 + 1}{\gamma/2 - \tan 0.75'} \qquad (2)$$

Moreover, in order to make sure the existence of MC, at least two rays should enter into the eye. So the lateral distance $t$ between two rays in front of the eye should be less than the pupil diameter ($d_{pup} = 4mm$ as usual), yielding,

$$t = z_P \gamma + h \leq d_{pup} \tag{3}$$

Eqs. (2) and (3) simultaneously determine three different areas of the HS. In the MC area, both of the equations are satisfied. While the whole scene is located within this area, it has compatible MC and binocular cues, and the true-3D imaging is observed. If only Eq. (3) is satisfied, a portion of the 3D scene is located out of the MC area and falls into the VMI area. The eye perceives multiple parallax images due to the wide separations between these images. Furthermore, if Eq. (3) is not satisfied, the objects are located in the LI area. Rays entering the eye are not sufficient and a distinct (sometimes discrete) scene is observed wherever the mono-eye focuses. There is no MC in the reconstruction. Therefore, both Eqs. (2) and (3) are the key conditions to *quantitatively* evaluate the MC in the HS.

Fig. 2(a) shows the three kinds of areas of a HS. The MC area (blue area) is divided into two parts by the LI area (dotted area), because less than two rays enter the mono-eye when the object is sufficiently close to the HS. Even the object is far from the LI area, it still has the chance to fall into the VMI area (white area). Figs. 2(b) and 2(c) show that the LI area can be diminished by reducing $d_H$ and $h$ respectively since more rays enter the mono-eye. In principle, the LI area vanishes as $h$ goes to zero. However, extremely small $h$ is challenging in the HS fabrication. In addition, the space-bandwidth product of a hogel should be at least twice as that of the elemental image, according to the sampling theorem. Therefore, appropriate parameters (e.g. the scene location, the HS location and its pixel) can be chosen in order to fabricate HS with better MC effect.

In order to demonstrate the MC in the HSs, a method based on the computational integral imaging is developed to simulate what the mono-eye can see through the HS. However, the previous formulas [4-5] are not suitable in our case since the mono-eye cannot see the entire elemental image. So a spatial-dependent pupil function $P_{ij}$ is introduced [see Fig. 3(a)], i.e.,

$$P_{ij}(x,y,z_F) = circ\left(\sqrt{(x-x_{oi})^2 + (y-y_{oj})^2}/a\right) = \begin{cases} 1, & (x-x_{oi})^2 + (y-y_{oj})^2 < a \\ 0, & others \end{cases}. \quad (4)$$

Here $a = (z_F - d_H)d_{pup}/2d_H$, $x_{oi} = x_e + z_F(x_{ui} - x_e)/d_H$, $y_{oj} = y_e + z_F(y_{uj} - y_e)/d_H$. $(x_e, y_e)$ and $(x_{ui}, y_{uj})$ are the central coordinates of the eye pupil and the $(i,j)$-th hogel, respectively. $(x,y)$ is the Cartesian coordinate of the reconstructed plane. Then the intensity on the retina yields,

$$I(x,y,z_F) = \frac{1}{N_s(x,y,z_F)} \sum_{i=1}^{N_x} \sum_{j=1}^{N_y} M\left[E_{ij}(x,y)\right] P_{ij}(x,y,z_F), \quad (5)$$

where $E_{ij}$ is the unmagnified elemental image and $M[\cdot]$ is the magnified elemental image, $N_s$ is the superposition number of each pixel for the magnified elemental image in the pupil. $N_x$ ($N_y$) is the number of elemental images which enter the eye in the x (y) direction. Here, the elemental images are generated by the ray-tracing technique from a 3D triangular-mesh models that can be obtained by the image-based modeling algorithm [13] or by a virtual computer-graphic model.

The 3D scene in our simulation contains five chesses. The king in front (the bishop in back) is at $z_F = 338(403)$ mm. The parameters of the HS are the same as those in Fig. 2(b), so that the whole scene falls into the MC area [light blue region in Fig. 2(b)]. The simulation results that focus on the front king [Fig. 3(b)] and the rear bishop [Fig. 3(c)] are presented. The bishop becomes defocus when the king is focused, and vice versa. This defocus effect further enhances the depth sensation and the correct MC effect. However, when the eye focuses into the VMI area, the VMI effect appears instead of the blur effect [see e.g. the horizontal edge line on the neck of the king becomes two visible lines in Fig. 3(d)]. Note that such effect can be emphasized if there are high-spatial-frequency textures on the object. As a result, if the scene depth is so large that some objects are located out of the MC area, the VMI effect makes the reconstructed image faked. In addition, Fig. 3(e) shows the discrete patterns in the reconstruction because the scene falls into the LI area.

The experiment is carried out to confirm MC area. Fig. 4(a) shows the HS printing system that is used to fabricate the full-parallax HS. A diffuser is placed on the front focal plane of the lens. The HS is close to a mask with a $0.5\text{mm} \times 0.5\text{mm}$ pinhole. The laser beam (532 nm), the spatial light modulator (SLM) and the step motor are controlled by a computer simultaneously. In the recording process, a serial elemental image is uploaded sequentially onto the SLM, while the HS is translated step by step. The same scene and parameters as those in Fig. 3(b) are used. The complete HS is $50\text{mm} \times 50\text{mm}$ with $100 \times 100$ hogels. A Nikon D6 digital camera is used to take photos focusing on different depths. A 4mm diameter pinhole was adhered to the camera to simulate the eye pupil. Two representative pictures are shown in Figs. 4(b) and 4(c) with the focal length of 340 and 400mm, which correspond to the cases in Figs. 3(b) and 3(c). The sharp edge of the front king appears when it is in focus, and vice versa. As the focus length varies, the in/out-of focus effect changes smoothly and there is no VMI or discrete patterns.

In conclusion, the MC in the HSs are studied. A quantitative condition is derived to identify the MC area. It enables us to choose parameters (e.g. the scene location, the HS location and its pixel), to ensure the reconstruction has correct MC and becomes true-3D imaging. A pupil-function integral imaging algorithm, as well as an experimental full-parallax HS, is used to confirm the theoretical predictions. In addition, our integral imaging algorithm is also a potential evaluation method in other multi-parallax 3D displays.

This work is supported by the NSFC (10804131, 10874250, and 10674183) and the FRFCU (2009300003161450).

**Figure captions**

Fig. 1. (Color online) Images on retina when the eye focuses on different distance planes through (a) a hologram and (b) a HS. (c) Schematic diagram for the quantitative condition derivation.

Fig. 2. (Color online) The MC area (blue), the LI area (dotted symbols), and the VMI area (white) of the HS with (a) $h = 0.5$mm, $d_H = 500$mm, (b) $h = 0.5$mm, $d_H = 230$mm, and (c) $h = 0.3$mm, $d_H = 500$mm. The light blue area in (b) corresponds to the MC area of the 3D scene used in Fig. 3(b).

Fig. 3. (Color online) (a) Schematic diagram on the pupil-function integral imaging algorithm. Numerical results when the scene is located (b)-(c) in the MC area [focusing (b) on the front king and (c) on the rear bishop], (d) in the VMI area ($z_F = 2000$mm), and (e) in the LI area ($z_F = 550$mm). (b)-(d) share the same 3D scene (placed within $z_p = 338 \sim 403$mm) and the same HS ($h = 0.5$mm, $d_H = 230$mm). The parameters of (e) are $z_p = 550 \sim 570$mm, $h = 0.5$mm, and $d_H = 500$mm.

Fig. 4. (Color online) (a) Experimental setup of the HS printing system. Optical reconstruction images when the mono-eye focuses on (b) the front king and (c) the rear bishop. The parameters of 3D scene and the HS are the same as those in Fig. 3(b).

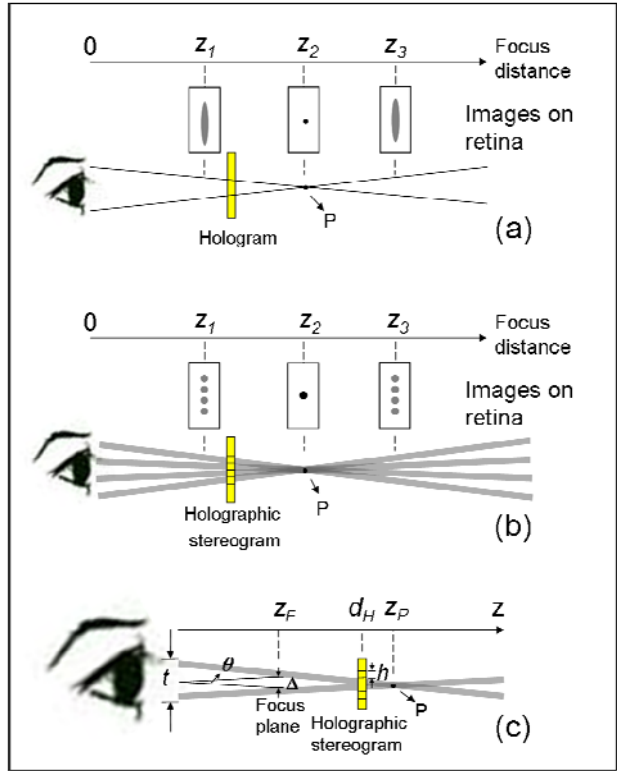

**Fig. 1**

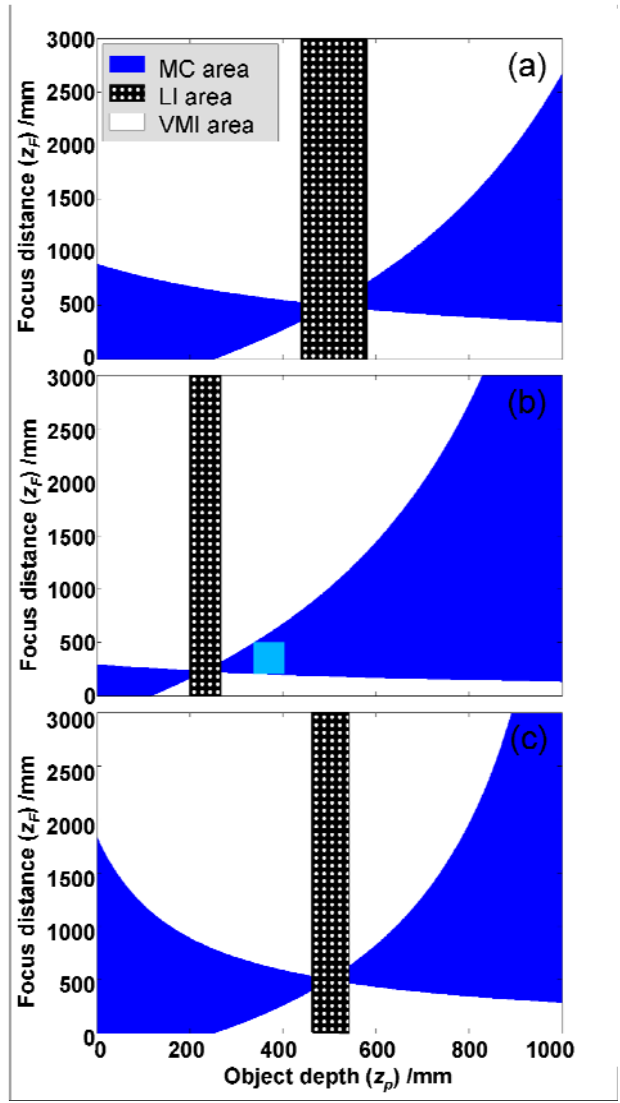

**Fig. 2**

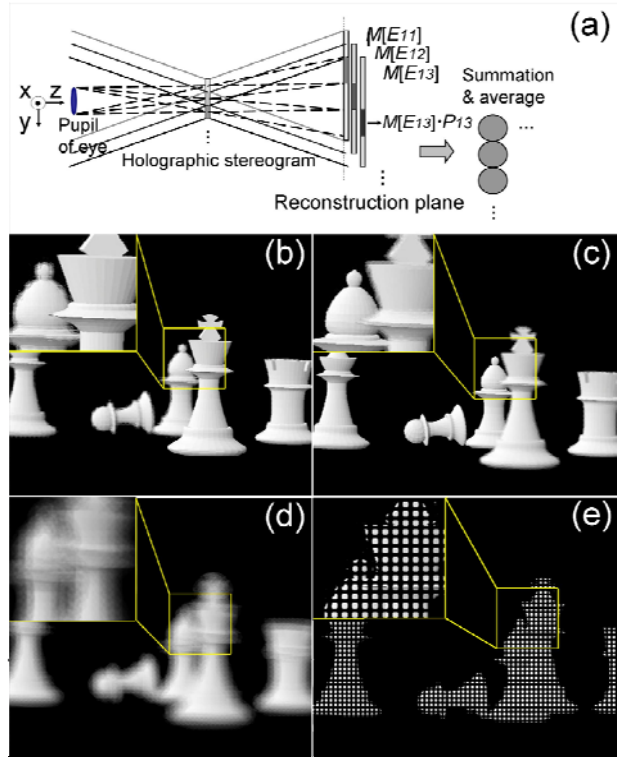

**Fig. 3**

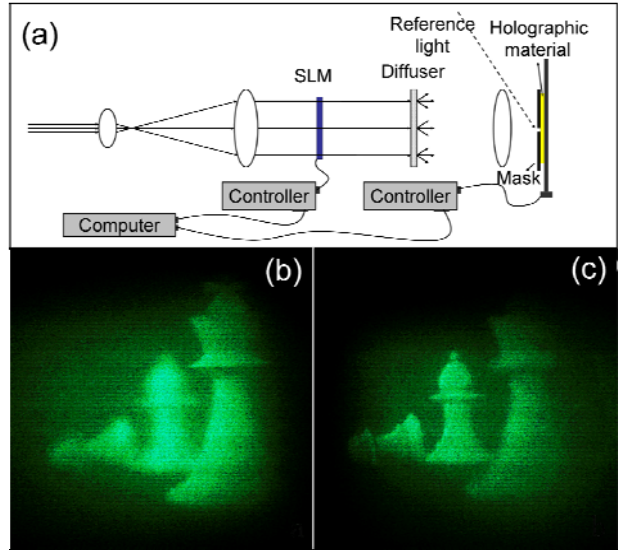

**Fig. 4**